\documentclass{appolb}
\pdfoutput=1
\usepackage[usenames,dvipsnames]{color}
\definecolor{darkblue}{RGB}{0,0,196}
\usepackage[colorlinks=true,linkcolor=darkblue,citecolor=darkblue,urlcolor=darkblue]{hyperref}
\usepackage{graphicx}
\usepackage{makecell}
\usepackage{bm}
\usepackage{amsmath}
\usepackage{amssymb}
\usepackage{sidecap}

\def\pt{p_\perp}

\begin{document}
\title{Thermalization and hydrodynamization in the
color-flux-tube model
\thanks{Talk presented by Radoslaw Ryblewski at \textit{Excited QCD 2016}, 6-12 March 2016, Costa da Caparica, Portugal.}
}
\author{Radoslaw Ryblewski
\address{The H. Niewodnicza\'nski Institute of Nuclear Physics, Polish Academy of Sciences, PL-31342 Krak\'ow, Poland \\~}
}

\maketitle
\begin{abstract}
The study of transverse-momentum spectra of quarks and gluons produced by the  color electric flux tube decaying through the Schwinger tunneling mechanism is reviewed. The hints for a fast hydrodynamization in the ultra-relativistic heavy-ion collisions are found.
\end{abstract}
\PACS{11.15Bt, 04.25.Nx, 11.10Wx, 12.38Mh}
 %
\section{Introduction}
\label{sec:intro}
%
\par It is now more than decade that the properties of the new state of matter called quark-gluon plasma (QGP) are studied with the ultra-relativistic heavy-ion collisions at RHIC and the LHC. Among many expected results there are also a few outstanding surprises. In particular, broad analysis of the particle correlations \cite{Bozek:2012en,Bozek:2015tca} in the experimental data revealed that, in contrast to the earlier expectations based on the asymptotic freedom property of quantum chromodynamics (QCD), the QGP is a strongly coupled system with the lowest viscosity in Nature.  Moreover, it is surprisingly well described within the effective framework of relativistic dissipative fluid dynamics. Unfortunately, predictive power of the latter, required for the ongoing precision studies, is highly limited due to the requirement of the precise knowledge of its initial conditions. In particular, one has only qualitative information on the amplitude of the viscous corrections just after the collision, which, apart from resulting in possible quantitative differences in the predicted results, may question applicability of the viscous fluid dynamics in such modelling in general \cite{Florkowski:2013lya,Denicol:2014mca,Jaiswal:2014isa,Tinti:2015xwa,
Nopoush:2015yga}. Thus, the study of the approach of the created QGP toward the local thermal equilibrium (\textit{thermalization}) is nowadays of highest interest for the field of heavy-ion physics \cite{Mrowczynski:1994xv,Rebhan:2008uj,Rebhan:2005re,
Gelis:2013rba,Berges:2013eia,Kurkela:2015qoa,Chesler:2010bi,
CaronHuot:2011dr,Heller:2011ju,vanderSchee:2013pia,Ryblewski:2013eja,
Jankowski:2014lna,Ruggieri:2015yea}.

\par In this proceedings contribution we briefly review our main results of the study of the QGP thermalization rate within the framework of the color-flux-tube model proposed in Ref.~\cite{Ryblewski:2013eja}. Based on the detailed analysis of the transverse-momentum spectra of the produced partons \cite{Ryblewski:2015psh} we show that it is difficult to fully thermalize the system within a short time ($\tau\lesssim 1$ fm) unless the shear viscosity to entropy density ratio is that of AdS/CFT lower bound, $\bar{\eta}_{\rm AdS} = 1/(4 \pi)$ \cite{Kovtun:2004de}. However, at the same time, for $\bar{\eta}\lesssim 3\, \bar{\eta}_{\rm AdS}$ the observed deviations from equilibrium state are quite well described within dissipative corrections of viscous hydrodynamics. The latter observation confirms the fast hydrodynamization of the QGP first proposed in Ref.~\cite{Heller:2011ju}.
%
\section{Color-flux-tube model for early stages of heavy-ion collisions}
%
Based on the saturation physics \cite{Lappi:2006fp}, due to the ultra-relativistic energies, most of particles produced in midrapidity of heavy-ion collisions originate from the decay of purely longitudinal color fields spanned by the receding color-charged nuclei. In  Refs.~\cite{Ryblewski:2013eja,Ryblewski:2015psh} it was proposed to describe this physics situation within the simplified model of initial color electric flux tube \cite{Casher:1978wy,Bialas:1984wv} treated in the Abelian dominance approximation \cite{Bialas:1986mt,Bialas:1987en} which subsequently decays into quarks and gluons according to the Schwinger tunneling mechanism \cite{Schwinger:1951nm}. Within this framework phase-space densities of quarks, antiquarks and charged gluons are described by the following kinetic Boltzmann-Vlasov-type equations
\begin{eqnarray}
\left( p^\mu \partial_\mu + g\,{\mbox{\boldmath $\epsilon$}}_i \cdot 
{\bf F}_{}^{\mu \nu } p_\nu \partial_\mu^p\right) f_{if}(x,p)& =& \frac{dN_{if}}{d\Gamma_{\rm inv} }+ C(f_{if}),  
\label{kineq}\nonumber\\
\left( p^\mu \partial_\mu - g\,{\mbox{\boldmath $\epsilon$}}_i \cdot 
{\bf F}_{}^{\mu \nu } p_\nu \partial_\mu^p\right) \bar{f}_{if}(x,p) &=& \frac{dN_{if}}{d\Gamma_{\rm inv} } + C(\bar{f}_{if}),  
\label{kineaq}\\
\hspace{-0.12cm}\left( p^\mu \partial _\mu + g\,{\mbox{\boldmath $\eta$}}_{ij} \cdot 
{\bf F}_{}^{\mu \nu } p_\nu \partial_\mu^p\right) \widetilde{{f}}_{ij}(x,p) &= &
\frac{d\widetilde{N}_{ij}}{d\Gamma_{\rm inv} }+ C(\widetilde{f}_{ij}), \label{kineg}\nonumber
\end{eqnarray}
respectively. In Eqs.~(\ref{kineaq}) partons couple to the mean color field ${\bf F}^{\mu\nu}=\left(F^{\mu\nu}_{(3)},F^{\mu\nu}_{(8)} \right)$ through the charges $\mbox{\boldmath $\epsilon$}_{i}$ (for quarks), $-\mbox{\boldmath $\epsilon$}_{i}$ (for antiquarks), and ${\mbox{\boldmath $\eta$}}_{ij}={\mbox{\boldmath $\epsilon$}}_{i}-{\mbox{\boldmath $\epsilon$}}_{j}$ (for gluons) \cite{Huang:1982ik}, where the color indices $i,j$ run from 1 to 3. The first terms on the right hand side of Eqs.~(\ref{kineaq}) denote the particle production due to the Schwinger tunnelling \cite{Ryblewski:2013eja,Schwinger:1951nm}. In addition to the usual treatment we include the subsequent collisions between produced particles through the collisional kernels treated in the relaxation time approximation (RTA) \cite{Bhatnagar:1954zz} 
$C(f)=p\cdot u \left(f_{\rm eq} -f\right)/\tau_{\rm eq}$
where the relaxation time   is expressed according to the Anderson-Witting formula $\tau_{\rm eq} = 5 \bar{\eta}/ T$ \cite{Anderson:1974}.
\par In the case of boost-invariant and transversely homogeneous system, also called the Bjorken symmetry \cite{Bjorken:1982qr}, the solutions of Eqs.~(\ref{kineaq}) depend only on the proper time, $\tau = \sqrt{t^2 - z^2}$, transverse momentum, $\pt$, and the boost-invariant variable, $w   =   t p_\parallel - z E$, i.e., $f=f(\tau,w,\pt)$, and may be found analytically \cite{Ryblewski:2013eja}.
%
\section{Transverse-momentum spectra}
%
In order to study approach of the system toward the local thermal equilibrium one typically analyses the deviations of pressures from the equilibrium pressure, see Ref.~\cite{Ryblewski:2013eja}. 
Alternatively, one can also study the $\pt$ spectra of the produced partons, and their approach toward the equilibrium spectrum \cite{Ryblewski:2015psh}. For that purpose we use Cooper-Frye formula \cite{Cooper:1974mv}, which for the Bjorken symmetry \cite{Bjorken:1982qr} considered herein reduces to the following form
\begin{eqnarray}
\frac{dN}{dy\, d^2p_\perp } =\pi R_\perp^2 \int\limits_{-\infty}^{+\infty} dw 
\,f(\tau,w,p_\perp),
\label{cfbi}
\end{eqnarray}
where $R_\perp$ arbitrarily sets the transverse size in such a way that the integral (\ref{cfbi}) gives the production per unit area.
 
We assume that the phase-space distribution $f$ in Eq.~(\ref{cfbi}) is given by the sum of the quark, antiquark and gluon distribution functions satisfying transport equations (\ref{kineaq}),
\begin{eqnarray}
 f  =  \sum_f^{N_f} \sum_i^3  \left(f_{if}  +   \bar{f}_{if} \right) 
+\sum_{i\neq j=1}^3  \widetilde{f}_{ij}, 
\label{f}
\end{eqnarray}
where $N_f =2$ is the number of flavours. For the results presented herein we use the solutions from Ref.~\cite{Ryblewski:2013eja} obtained for the initial field strength corresponding to the LHC case.
\section{Results}
%
\par
\begin{figure}[t]
\begin{center}
\includegraphics[width=0.49 \textwidth]{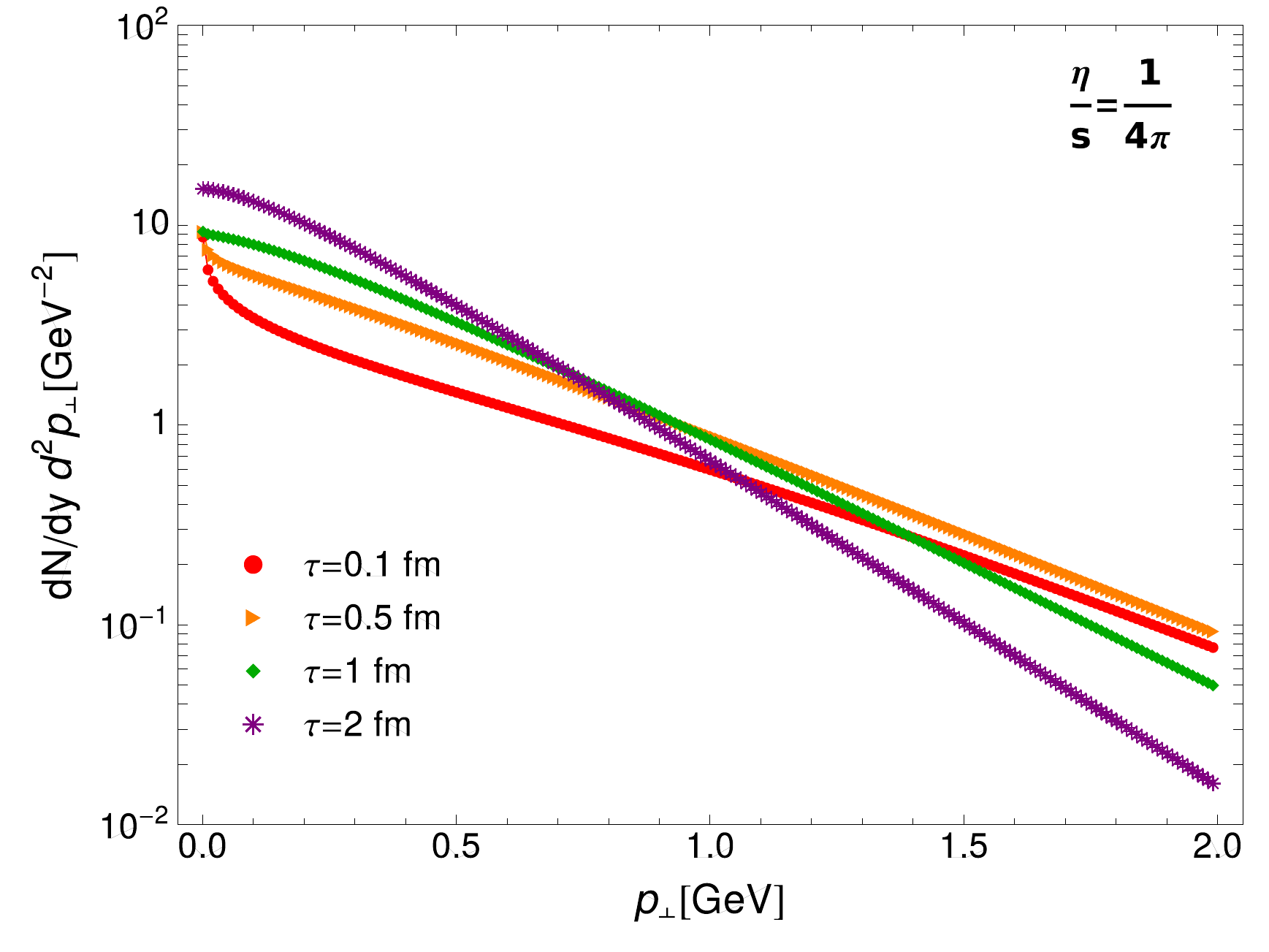} \includegraphics[width=0.49 \textwidth]{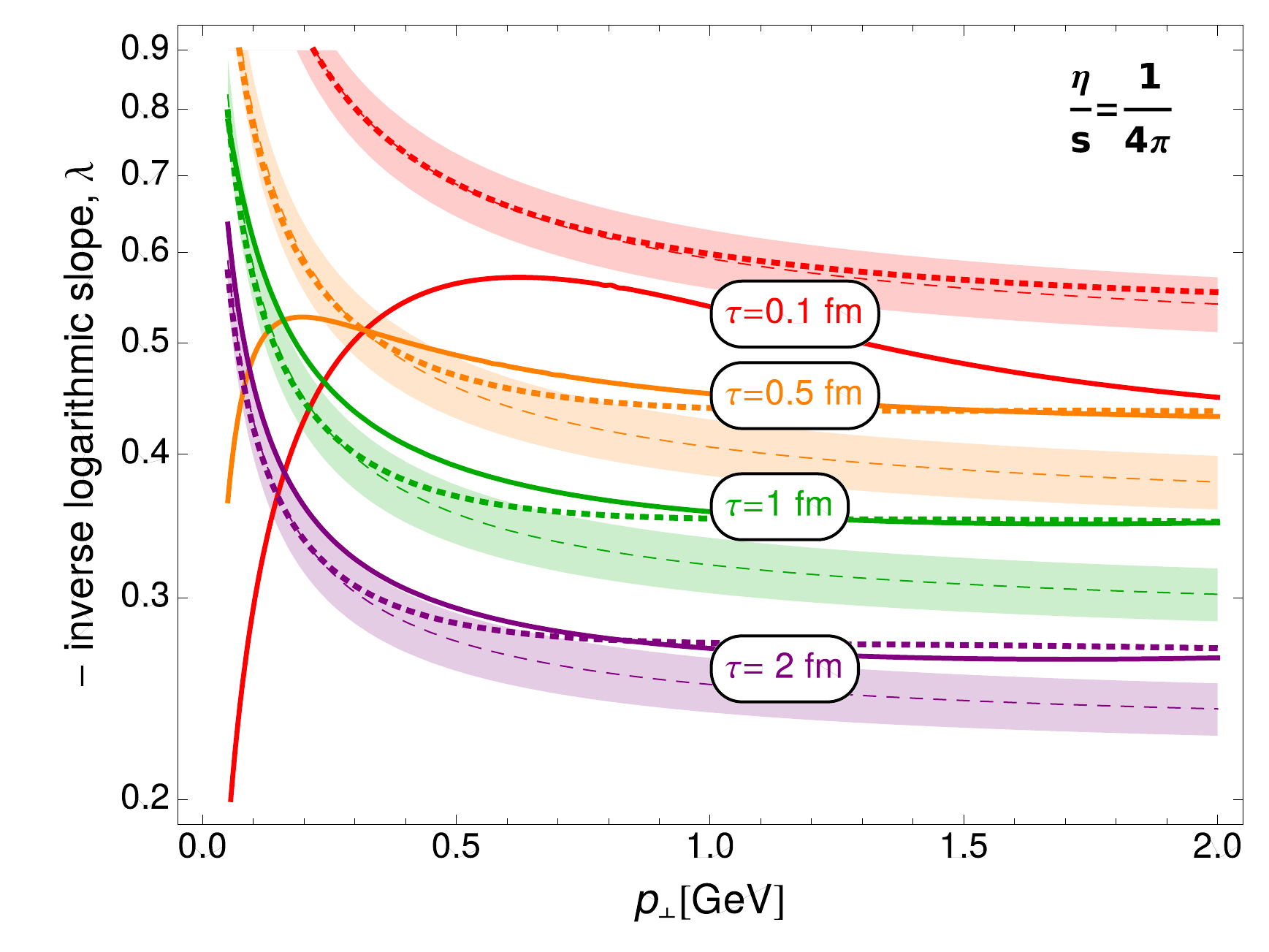}\\

\includegraphics[width=0.49 \textwidth]{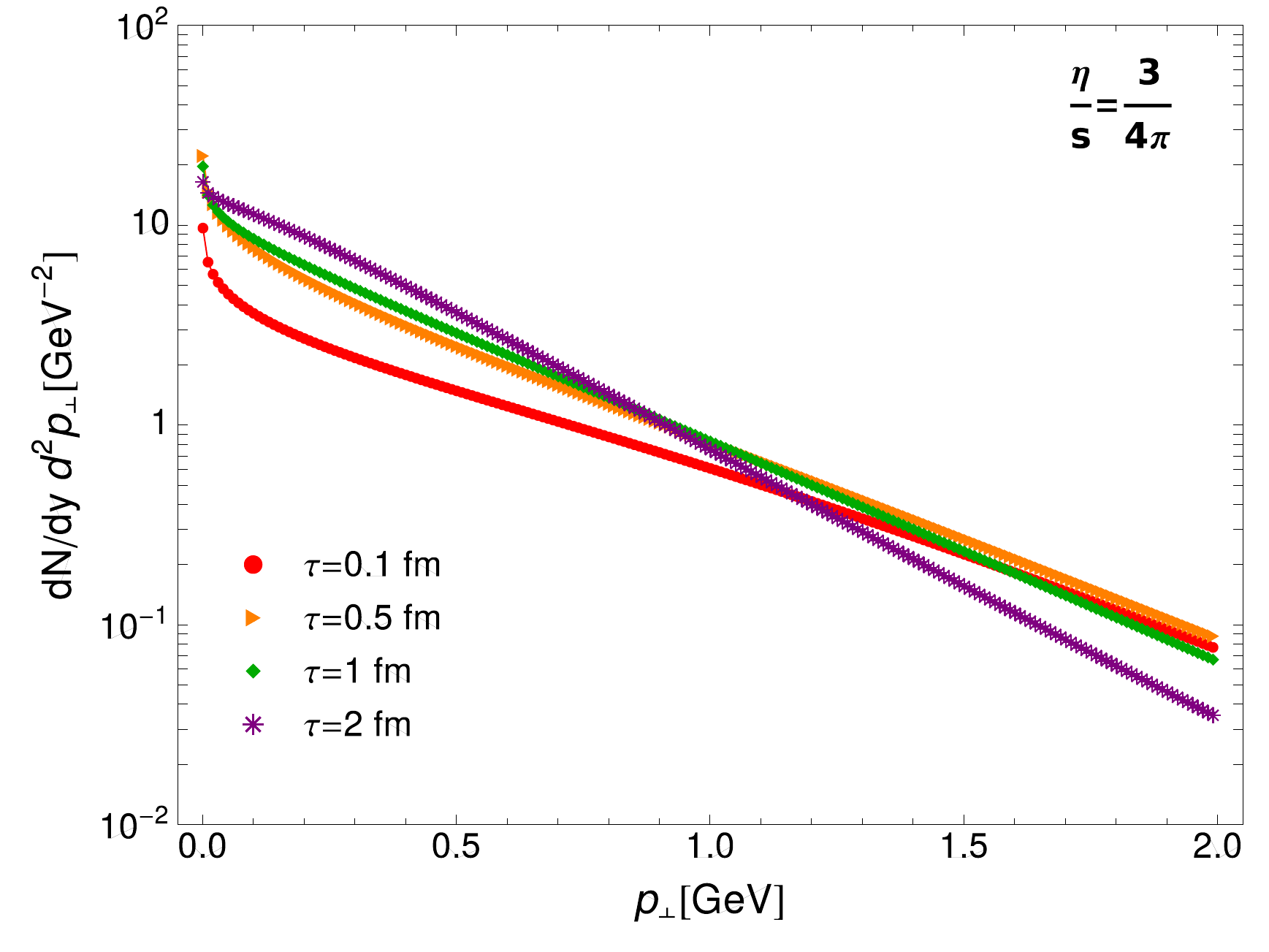}
\includegraphics[width=0.49 \textwidth]{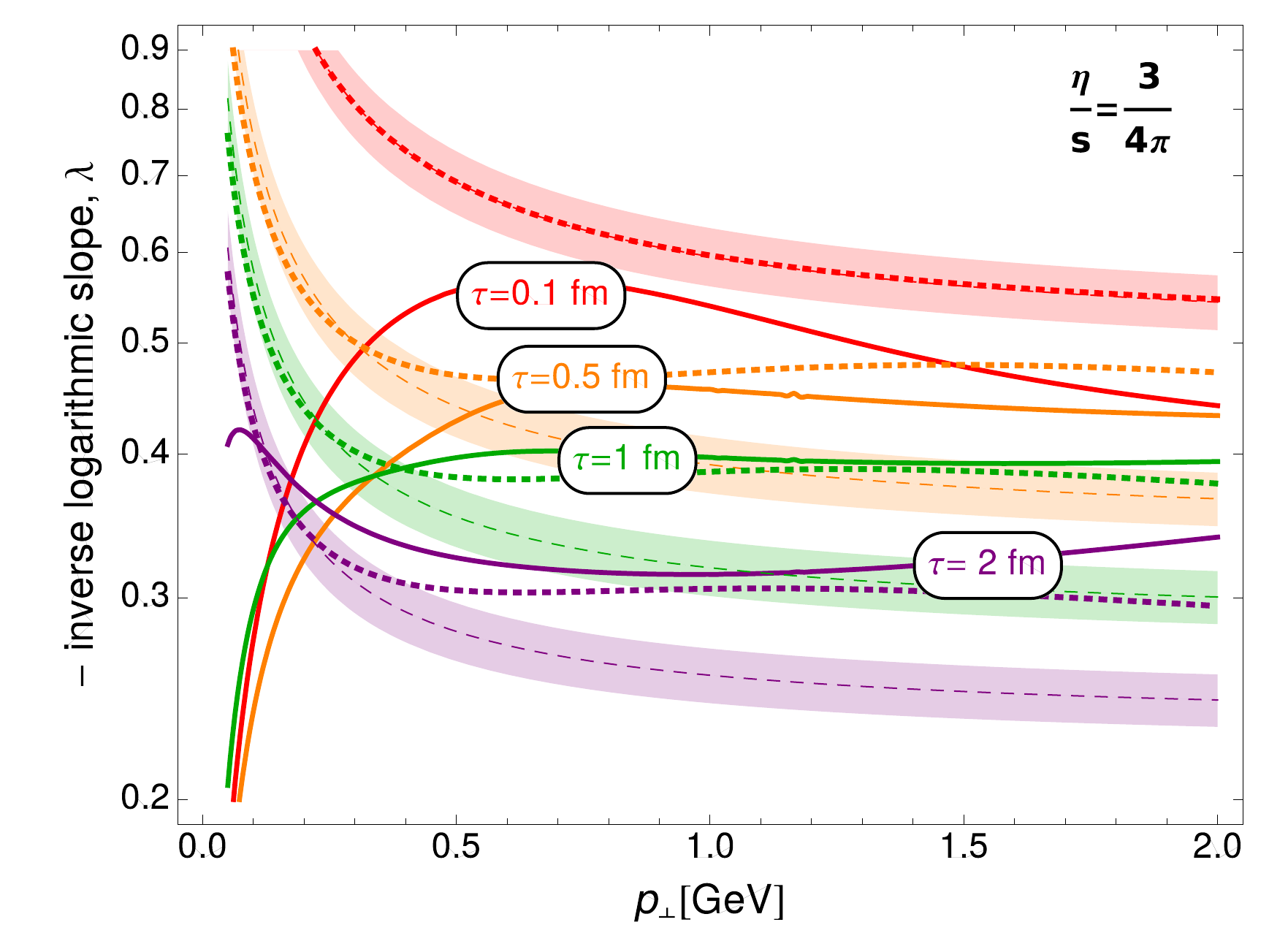}\\
\vspace{-0.2cm}
\end{center}
\caption{
The $\pt$ spectra of partons (left panels) and their corresponding inverse logarithmic slopes (right panels) for ${\bar \eta}= \bar{\eta}_{\rm AdS}$ (top) and ${\bar \eta} = 3 \,\bar{\eta}_{\rm AdS}$ (bottom), and various freeze-out proper times. The model predictions slopes $\lambda$ (solid lines) are compared to the slopes of the corresponding equilibrium spectrum $\lambda_{\rm eq}$ (dashed lines) and the viscous fluid spectrum $\lambda_{\rm visc}$ (dotted lines). 
}
\label{sp1}
\end{figure}

 In the left panels of Fig.~\ref{sp1} we present the partonic $\pt$ spectra obtained using Eqs.~(\ref{cfbi}) and (\ref{f}) for two values of the shear viscosity to entropy density ratio, $\bar{\eta} = \bar{\eta}_{\rm AdS}$ (top) and $\bar{\eta} = 3\, \bar{\eta}_{\rm AdS}$ (bottom), and various freeze-out proper times. We observe that, except for the low-$\pt$ range, in all cases the spectra have approximately thermal-like (exponential) shape. This property, sometimes called \textit{apparent thermalization}, is, however, mainly related to the specific production mechanism of partons, which in our case is the Schwinger tunnelling in the oscillating color field \cite{Bialas:1999zg,Florkowski:2003mm}. 
\par In order to judge if the thermalization is really achieved we propose to study the corresponding \textit{negative inverse logarithmic slopes}, $\lambda$, of the spectra,
\begin{equation}
\lambda = -\left[\frac{d}{d\pt} \ln\left(\frac{dN}{dy\, d^2p_\perp }\right)\right]^{-1}.\label{slope}
\end{equation}
%
%
%
%
In the right panels of Fig.~\ref{sp1} we compare the slopes $\lambda$ of the spectra from the left panels with the corresponding slopes of the spectra of the viscous fluid,
\begin{eqnarray}
\lambda_{\rm visc}  &=& T \frac{((a \hat{p}_\perp)^{-1}+ \hat{p}_\perp) K_1(\hat{p}_\perp) -2 K_2(\hat{p}_\perp)}{((a \hat{p}_\perp)^{-1}+ \hat{p}_\perp) K_0(\hat{p}_\perp) -4 K_1(\hat{p}_\perp)},
\label{lvisc}
\end{eqnarray}
having the same energy density $\varepsilon(\tau)$. In Eq.~(\ref{lvisc}) $\hat{p}_\perp=\pt/T$ and $a = (P_T - P_L)/(8\,\varepsilon)$, with $P_L$ and $P_T$ being the longitudinal and transverse pressure, respectively. In the special case of the system in local thermal equlibrium, $P_T = P_L$, the formula (\ref{lvisc}) reduces to the one for the perfect fluid,
\begin{eqnarray}
 \lambda_{\rm eq} = \lim_{a\to0} \lambda_{\rm visc} = T K_1 (\pt/T)/ K_0 (\hat{p}_\perp).
 \label{leq}
\end{eqnarray}
From the Fig.~\ref{sp1} we observe that, although the spectra look approximately thermal for all $\tau$, only in the case of $\bar{\eta} = \bar{\eta}_{\rm AdS}$ the slopes $\lambda$ agree with the equilibrium ones $\lambda_{\rm eq}$ within a short proper-time ($\tau\lesssim 1$ fm). This suggest that the system undergo fast thermalization only in the case of maximal possible coupling. On the other hand, the discrepancies at medium and large $\pt$ observed for $\bar{\eta} = 3\,\bar{\eta}_{\rm AdS}$ (bottom panel) are very well addressed within the dissipative corrections of the viscous fluid dynamics, see $\lambda_{\rm visc}$. It means that although the fast thermalization seems to be limited to $\bar{\eta} = \bar{\eta}_{\rm AdS}$, the system may still undergo fast hydrodynamization if $\bar{\eta}\lesssim 3\,\bar{\eta}_{\rm AdS}$.
%
\section{Conclusions}
The thermalization rate of the quark-gluon plasma produced from the initial color electric flux tubes through the Schwiner mechanism is studied. Once produced the partons are interacting with each other according to the collisional kernels treated in RTA. The detailed comparison of the resulting transverse-momentum spectra of partons with the viscous hydrodynamics ones is done. Obtained results suggest that, although the early thermalization may be limited to strongly coupled systems ($\bar{\eta} =  \bar{\eta}_{\rm AdS}$), as long as $\bar{\eta}\lesssim 3\,\bar{\eta}_{\rm AdS}$ the early hydrodynamization of the plasma in such reactions may be achieved. 
\section*{Acknowledgments}
R.R. was supported by Polish National Science Center Grant No. DEC-2012/07/D/ST2/02125.


\begin{thebibliography}{100}
\expandafter\ifx\csname url\endcsname\relax \def\url#1{{\tt #1}}\fi
\expandafter\ifx\csname urlprefix\endcsname\relax\def\urlprefix{URL}\fi
\providecommand{\eprint}[2][]{\url{#2}}

\bibitem{Bozek:2012en} 
  P.~Bozek and W.~Broniowski,
   {\em Phys.\ Rev.\ Lett.}   {\bf 109}, 062301 (2012).
  
\bibitem{Bozek:2015tca} 
  P.~Bozek, W.~Broniowski and A.~Olszewski,
  {\em Phys.\ Rev.} {\bf C92}, no. 5, 054913 (2015).
  
\bibitem{Florkowski:2013lya} 
  W.~Florkowski, R.~Ryblewski and M.~Strickland,
  {\em Phys.\ Rev.} {\bf C88}, 024903 (2013).
  
\bibitem{Denicol:2014mca} 
  G.~S.~Denicol, W.~Florkowski, R.~Ryblewski and M.~Strickland,
   {\em Phys.\ Rev.}  {\bf C90}, no. 4, 044905 (2014).

\bibitem{Jaiswal:2014isa} 
  A.~Jaiswal, R.~Ryblewski and M.~Strickland,
  {\em Phys.\ Rev.} {\bf C90}, no. 4, 044908 (2014).
 
\bibitem{Nopoush:2015yga} 
  M.~Nopoush, M.~Strickland, R.~Ryblewski, D.~Bazow, U.~Heinz and M.~Martinez,
  {\em Phys.\ Rev.}  {\bf C92}, no. 4, 044912 (2015).
    
    
\bibitem{Tinti:2015xwa} 
  L.~Tinti,
  arXiv:1506.07164 [hep-ph].
   

\bibitem{Mrowczynski:1994xv} 
  S.~Mrowczynski,
  {\em  Phys.\ Rev.} {\bf C49}, 2191 (1994).

\bibitem{Rebhan:2008uj} 
  A.~Rebhan, M.~Strickland and M.~Attems,
  {\em Phys.\ Rev.} {\bf D78}, 045023 (2008).

\bibitem{Rebhan:2005re} 
  A.~Rebhan, P.~Romatschke and M.~Strickland,
  {\em JHEP} {\bf 0509}, 041 (2005).
%
%
\bibitem{Gelis:2013rba} 
  T.~Epelbaum and F.~Gelis,
  {\em Phys.\ Rev.\ Lett.} {\bf 111}, 232301 (2013).

\bibitem{Berges:2013eia} 
  J.~Berges, K.~Boguslavski, S.~Schlichting and R.~Venugopalan,
  {\em Phys.\ Rev.} {\bf D89}, no. 7, 074011 (2014).
 
\bibitem{Kurkela:2015qoa} 
  A.~Kurkela and Y.~Zhu,
  {\em Phys.\ Rev.\ Lett.} {\bf 115}, no. 18, 182301 (2015).
  
\bibitem{Chesler:2010bi} 
  P.~M.~Chesler and L.~G.~Yaffe,
  {\em Phys.\ Rev.\ Lett.} {\bf 106}, 021601 (2011).
  
\bibitem{CaronHuot:2011dr} 
  S.~Caron-Huot, P.~M.~Chesler and D.~Teaney,
  {\em Phys.\ Rev.} {\bf D84}, 026012 (2011).
   
\bibitem{Heller:2011ju} 
  M.~P.~Heller, R.~A.~Janik and P.~Witaszczyk,
  {\em Phys.\ Rev.\ Lett.}  {\bf 108}, 201602 (2012).

\bibitem{vanderSchee:2013pia} 
  W.~van der Schee, P.~Romatschke and S.~Pratt,
  {\em  Phys.\ Rev.\ Lett.}  {\bf 111}, no. 22, 222302 (2013).
  
\bibitem{Ryblewski:2013eja} 
  R.~Ryblewski and W.~Florkowski,
  {\em Phys.\ Rev.} {\bf D88}, 034028 (2013).
  
\bibitem{Jankowski:2014lna} 
  J.~Jankowski, G.~Plewa and M.~Spalinski,
  {\em JHEP} {\bf 1412}, 105 (2014).

\bibitem{Ruggieri:2015yea} 
  M.~Ruggieri, A.~Puglisi, L.~Oliva, S.~Plumari, F.~Scardina and V.~Greco,
  {\em Phys.\ Rev.} {\bf C92}, 064904 (2015).
  
\bibitem{Ryblewski:2015psh} 
  R.~Ryblewski,
  arXiv:1512.04117 [nucl-th].
  
\bibitem{Kovtun:2004de} 
  P.~Kovtun, D.~T.~Son and A.~O.~Starinets,
  {\em Phys.\ Rev.\ Lett.}  {\bf 94}, 111601 (2005).

\bibitem{Lappi:2006fp} 
  T.~Lappi and L.~McLerran,
  {\em Nucl.\ Phys.} {\bf A772}, 200 (2006).
  
\bibitem{Casher:1978wy} 
  A.~Casher, H.~Neuberger and S.~Nussinov,
  {\em Phys.\ Rev.} {\bf D20}, 179 (1979).
  
\bibitem{Bialas:1984wv} 
  A.~Bialas and W.~Czyz,
  {\em Phys.\ Rev.} {\bf D30}, 2371 (1984).
  
\bibitem{Bialas:1986mt} 
  A.~Bialas and W.~Czyz,
  {\em Acta Phys.\ Polon.} {\bf B17}, 635 (1986).
 
\bibitem{Bialas:1987en} 
  A.~Bialas, W.~Czyz, A.~Dyrek and W.~Florkowski,
  {\em Nucl.\ Phys.} {\bf B296}, 611 (1988).
   
\bibitem{Schwinger:1951nm} 
  J.~S.~Schwinger,
  {\em Phys.\ Rev.}  {\bf 82}, 664 (1951).
  
\bibitem{Huang:1982ik} 
  K.~Huang,
  Singapore, Singapore: World Scientific (1992) 333p.
  
\bibitem{Bhatnagar:1954zz} 
  P.~L.~Bhatnagar, E.~P.~Gross and M.~Krook,
  {\em Phys.\ Rev.} {\bf 94}, 511 (1954).
  
\bibitem{Anderson:1974}
  J.~L.~Anderson and H.~R.~Witting, 
  {\em Physica}  {\bf 74}, 466–488 (1974).
  
\bibitem{Bjorken:1982qr} 
  J.~D.~Bjorken,
  {\em Phys.\ Rev.} {\bf D27}, 140 (1983).
  
\bibitem{Cooper:1974mv} 
  F.~Cooper and G.~Frye,
  {\em Phys.\ Rev.} {\bf D10}, 186 (1974).
 
\bibitem{Bialas:1999zg} 
  A.~Bialas,
  {\em Phys.\ Lett.} {\bf B466}, 301 (1999).
  
\bibitem{Florkowski:2003mm} 
  W.~Florkowski,
  {\em Acta Phys.\ Polon.} {\bf B35}, 799 (2004).
 

\end{thebibliography}
\end{document}